\documentclass[conference]{IEEEtran}
\IEEEoverridecommandlockouts                              

\usepackage[utf8]{inputenc}
\usepackage{bm}

\usepackage{graphicx} 
\usepackage{epsfig} 
\usepackage{amsmath} 
\usepackage{amssymb}  
\usepackage{enumitem}

\usepackage{hyperref}
\usepackage[dvipsnames]{xcolor}
\hypersetup{
	colorlinks=false,
	linkcolor=blue,
	filecolor=magenta,      
	urlcolor=cyan,
	pdftitle={Controller-Aware Dynamic Network Management for Industry 4.0},
	allbordercolors = OliveGreen,
	hidelinks = true
}
\makeatletter
\renewcommand*{\eqref}[1]{%
	\hyperref[{#1}]{\textup{\tagform@{\ref*{#1}}}}%
}
\makeatother

\title{\LARGE \bf
Controller-Aware Dynamic Network Management for Industry~4.0
}

\author{Efe C. Balta, Mohammad H. Mamduhi, John Lygeros, and Alisa Rupenyan
\thanks{This work is supported by NCCR Automation, funded by the Swiss National Science Foundation through Grant no. 180545.}
\thanks{E. C. Balta, M. H. Mamduhi, J. Lygeros, and A. Rupenyan are with the Automatic Control Laboratory, Department of Information Technology and Electrical Engineering, ETH Z\"urich, Switzerland.
        {\tt\small email: \{ebalta,mmamduhi,jlygeros,ralisa\}@ethz.ch}}%
}

\begin{document}

\maketitle
\thispagestyle{empty}
\pagestyle{empty}

\begin{abstract}

 In this paper, we consider a cyber-physical manufacturing system (CPMS) scenario containing physical components (robots, sensors, and actuators), operating in a digitally connected, constrained environment to perform industrial tasks. The CPMS has a centralized control plane with digital twins (DTs) of the physical resources, computational resources, and a network manager that allocates network resources.  
  Existing approaches for allocation of network resources are typically fixed with respect to controller-dependent run-time specifications, which may impact the performance of physical processes. 
We propose a dynamic network management framework, where the network resource allocation schemes are controller-aware. The information about the controllers of the physical resources is implemented at the DT level, and metrics, such as regret bounds, take the process performance measures into account. The proposed network management schemes optimize physical system performance by balancing the shared resources between the physical assets on the plant floor, and by considering their control requirements, providing a new perspective for dynamic resource allocation. A simulation study is provided to illustrate the performance of the proposed network management approaches and compare their efficiencies.

\end{abstract}

\section{Introduction}

Industrial Cyber-Physical Systems (ICPS) generally involve a layer of coupled mechanical/physical components, and a cyber layer that supports the interactions and data exchange between the components of the physical layer and additionally with the monitoring, remote supervision and maintenance units \cite{8558534}. Modern applications of ICPS require sensitive control functionalities across the physical layer, and a reliable and agile cyber layer enabling sustainable and coordinated interactions  \cite{Allgoewer2019,7778207}. Traditionally, the development of ICPS was somehow synonymous with the development of control architectures, e.g., the progress in deploying distributed control systems for process industries, or advancements in programmable logic controllers (PLC) for discrete manufacturing, while the communication was merely in the form of exchanging pre-determined control and actuation command signals via locally installed networks mainly with wired channels \cite{8840800}. 

The increasing use of the 5th-generation Internet (5G), together with the fast development of advanced computational methods and in-network computations (edge and fog computing) provide a plethora of opportunities for ICPS to move into a new level of autonomy, reconfigurability, and efficiency  \cite{8403588,8612449}. State-of-the-art networking technology can support a wide range of vertical industries within one physically installed network infrastructure, through its virtual and programmable features \cite{9200919,8931325}. 
Additionally, software-defined and virtual components, e.g., software-defined networks (SDN) enable demand-driven,  status-aware data networks that allow for customized quality-of-service optimized for the physical components in the network~\cite{8853247,Bonati2020}.

In the context of manufacturing, ICPS are often referred to as Cyber-Physical Manufacturing Systems (CPMS). Efficient network resource allocation among service-critical physical components is a challenging problem in a CPMS scenario, due to the time-varying resource requirements that are sensitive to the communication and computation needs \cite{8660405}. 
The novel networking features and capabilities have led to the development of more advanced and reconfigurable resource allocation architectures  \cite{8399557,59155,qamsane2019unified,kovalenko2022towards}. Learning and AI methods, software-defined resource orchestration, and virtual sub-networks have been proposed for automation and control of CPMS with efficient resource allocation
\cite{kovalenko2022towards,lopez2018software,9422344}. 
Hierarchical resource allocation mechanisms are popular for resource provisioning in industrial plants due to their flexibility in deployment and capability of being implemented in distributed fashion \cite{8311660,8553664}. Nevertheless, a deployable resource allocation mechanism that supports heterogeneous industrial operations \emph{in a controller-aware fashion} is not addressed in the literature.

The main goal of this paper is to provide a controller-aware and sustainable resource allocation and networking management for the state-of-the-art CPMS. 
We assume that the CPMS has a centralized control plane that has access to the plant floor data through a unified data interface, see, e.g.,~\cite{qamsane2019unified,lopez2018software}, and a pool of digital twins (DTs) that represent the physical plant floor through models, measurement data, and algorithmic intelligence~\cite{moyne2020requirements,qamsane2019unified,7976223}. The communication/computation requirements of the physical resources vary based on a number of factors, which are monitored by the responsible DTs. We propose a dynamic network management framework, where the controller-aware requirements and run-time measurements gathered through the DTs are used by the network manager for optimal network resource allocation, according to set requirements. We explore event-based reallocation of network resources, where the events are triggered based on controller performance-aware metrics, and online (continuous) reallocation scheme, and compare them with existing network allocation approaches. We demonstrate the efficiency of the proposed controller-aware approaches in a numerical case study and show that they greatly outperform the static network allocation methods without run-time data from the DTs.

As the remainder of this paper, Section \ref{SecII:model} describes the CPMS model and the problem formulation. We propose our controller-aware resource allocation architecture in Section \ref{SecIII:RA}. In Section \ref{SecIV:SA} the simulation results are demonstrated and the paper is concluded in Section \ref{SecV:Con}.

\section{CPMS Model and Problem Formulation}
\label{SecII:model}

\begin{figure}
	\centering
	\includegraphics[width=\columnwidth]{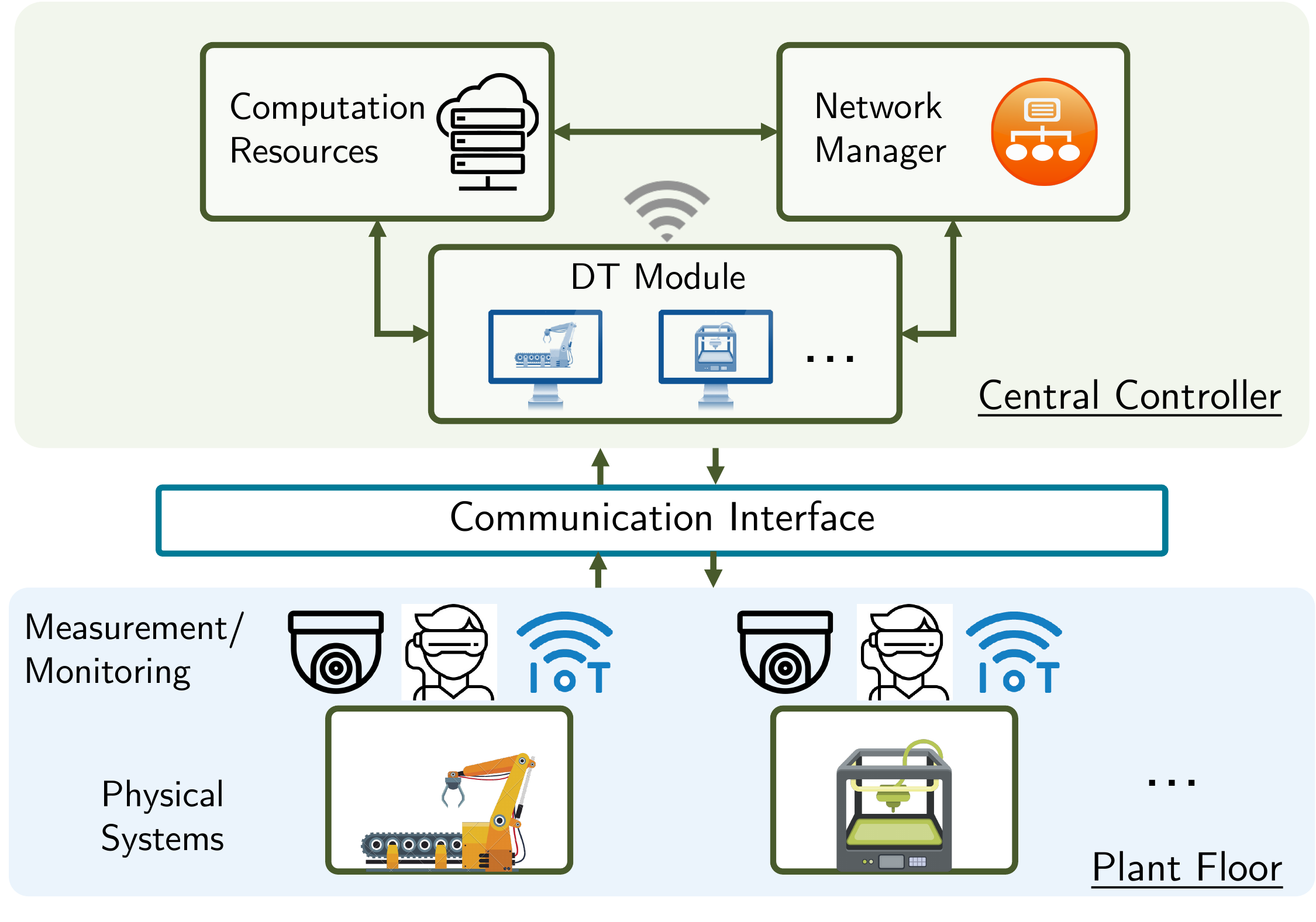}
	\caption{Overview of the CPMS setting considered in this work.}
	\label{fig:setting}
\end{figure}

We consider a CPMS scenario wherein multiple dynamically heterogeneous components are operating to perform given industrial tasks including sensing, monitoring, and actuation. The CPMS setting is schematically illustrated in Figure~\ref{fig:setting}. The physical systems make up the plant floor of the CPMS. Each physical system is equipped with various sensors, industrial internet-of-things devices, etc. A unified communication interface collects data from the heterogeneous devices from the plant floor, and makes it available to the software modules in the central controller, e.g., a southbound interface as proposed in~\cite{lopez2018software,qamsane2019unified}. The central controller has a DT module that includes the DTs, databases, and necessary application interfaces. 
The network manager utilizes the data from the DTs and the plant floor to access the computation resources and optimize the necessary allocation. The computational resources may be part of the centralized controller plane, or edge devices on the plant floor, distributed resources connected with 5G networks, etc. Using the network allocation, the DTs compute the control inputs and send them to the physical systems on the plant floor.

The DTs in the central controller are tasked with monitoring the physical assets and processes against task requirements, communicate measurements, metrics, and controller information with the network manager, and provide control actions to the plant floor. The network requirements of physical resources and processes are evaluated by the DTs and shared with the network manager, so that the network manager can optimally allocate resources in order to maximize performance.

Control inputs for each physical component are computed based on the simulated operations, the task requirements, and constraints, assuming that the required communication resources are provided by the network. Since network resources need to be shared among all physical components, there might be temporary mismatches between service requirements and network serviceability in terms of providing tailored services with exact characteristics such as latency or bandwidth. In addition, industrial plants might be exposed to disturbances and malfunction, resulting in potential discrepancies between the expected and real performance of each physical component. Similarly, changing operation characteristics due to the interactions between the physical resources, or reconfigurations on the plant floor may result in changes in required network resources.
The control action computed by the DT might be an actuation command, or a reference trajectory to be executed or tracked by the physical resource or process on the plant floor.

\section{Controller-aware Resource Allocation}
\label{SecIII:RA}
We start by considering static allocation problems and move toward online and event-triggered controller-aware resource allocation schemes; in the process we outline the role of DTs in our proposed framework.

\subsection{Static to Dynamic Allocation}
In the \emph{static allocation} setting, allocation is done offline, based on the expected requirements and each resource gets the predetermined allocation in run-time.
A generalized static allocation problem can be written as 
\begin{align}\label{eq:static_alloc}
	\min\limits_{a\in \mathcal{A}}~&\{ J(a, r) |  \sum a_i \leq b \},
\end{align}
where $\textstyle{a \in \mathbb{R}^{n_a}}$ is the allocation of each asset in the network for all times, $\mathcal{A}$ denotes constraints such as non-negativity, additional lower-bounds, or limits, and $b$ is the total available resource in the network, e.g., total bandwidth, computation capacity.  $J(a,r)$ is a cost function designed to ensure optimal allocation configurations in the network, which often penalizes deviations from target allocation values. Note that the inequality constraint may be exchanged with a simplex constraint, whenever applicable.

Resource allocation according to scheme in \eqref{eq:static_alloc} relies on the expected requirements of the physical assets, denoted by $\textstyle{r\in\mathbb{R}^{n_a}}$, which is not robust to disturbances and run-time changes in the requirements, and does not allow for easy reconfiguration. Therefore, while static allocation may be adequate in certain settings, an interconnected I4.0 system with dynamically changing plant floor environment requires higher robustness and flexibility.

Dynamic resource allocation schemes utilize run-time information about the plant floor and the physical assets to allocate the network resources based on the current-time requirements. Such applications are enabled through efficient communication protocols and increased edge intelligence on devices that allow them to have better cognition.
The dynamics allocation problem can be stated as
\begin{align}\label{eq:dynamic_alloc}
	\min\limits_{\bm{a}}~&\{J_N(\bm{a}, \bm{r}) | H \bm{a} \leq  \bm{b}\}, 
\end{align}
where $\textstyle{\bm{a} = \{a_{t_0}, \ldots,a_{t_{N-1}} \}}$ is a horizon of allocation for the assets,  $J_N(\bm{a}, \bm{r})$ denotes a cost function over the horizon of $N$, $\textstyle{\bm{r} = \{r_{t_0}, \ldots,r_{t_N} \}}$ denotes the forecast of requirements over the horizon, and $\textstyle{H \bm{a} \leq  \bm{b}}$ denotes input constraints and network capacity constraints over the horizon of $N$, including the set $\mathcal{A}$ in \eqref{eq:static_alloc}. Note that \eqref{eq:dynamic_alloc} is a multi time-step generalization of \eqref{eq:static_alloc}, and the planning horizon $N$ is problem specific. After the problem \eqref{eq:dynamic_alloc} is solved, the corresponding optimal allocation may be implemented until the end of a horizon, which may require a re-optimization. Alternatively, the problem can be solved in receding horizon, as we propose next.

\subsection{Online Dynamic Allocation}
To take the controller requirements and performance into account, we define additional constraints and cost terms denoting the controller performance under a given network allocation. In this setting, the cost function $J_N$ may utilize an online forecast $\textstyle{\bm{r}} $ that is updated via information collected from the plant floor in run-time. We assume here that this is a deterministic forecast, though uncertain forecasts can also be considered through robust~\cite{langson2004robust} and stochastic~\cite{mesbah2016stochastic} formulations can also be considered. The controller-aware resource allocation problem is defined as
\begin{subequations}
\label{eq:online_alloc}
\begin{align}
	\min\limits_{\bm{a}}~& \mathcal{I}(\bm{\gamma}) + \sum_{t = t_0}^{t_0+N-1} \ell(\xi_t,r_t,a_t) + \ell(\xi_{t_0+N},r_{t_0+N}), \\
	\text{s.t.:}~& H(\bm{r}) a \leq  \bm{b}(\bm{r}) + \bm{\gamma},\quad \xi_{t_0} = \hat{\xi} \\
	   ~&\xi_{t+1} = A \xi_t + B a_t,~ t = t_0,\ldots, t_0 +N-1, \label{eq:nw_dyn}
\end{align}
\end{subequations}
where $\textstyle{\bm{\xi} = \{\xi_{t_0}, \ldots,\xi_{t_N} \}}$ is the predicted state evolution for the network with the dynamics given in \eqref{eq:nw_dyn}, $\hat{\xi}$ is the current state of the network, and $\ell$ is a well-defined loss function on the difference between the reference and the state.
The variable $\bm{\gamma}$ is a slack variable for satisfying the constraints represented by $H(\bm{r})$ and $\bm{b}(\bm{r})$, with $\mathcal{I}(\bm{\gamma})$ representing the corresponding penalty function.

The constraints represent allocation input constraints, as well as additional constraints based on the reference $\bm{r}$, such as maximum allowable deviation, or minimum allowable allocation for a given physical resource. Additionally, the state of the system $\xi(t)$ may represent state variables that are not only the allocated network, etc., therefore the formulation of \eqref{eq:online_alloc} allows us to consider a general class of online network allocation problems.

We interpret the minimization problem in \eqref{eq:online_alloc} as a receding horizon control problem that accounts for the network dynamics through $A$ and $B$. For nonzero $A$ matrix, we have inertia of delays in the system, meaning that the network allocation for some of the resources cannot be arbitrarily changed between consecutive time-steps. Matrix $B$ represents input dynamics for the allocation, e.g., if there are proportional losses in the allocated network, or if there is efficiency involved in the allocation.

An online dynamic network allocation scheme obtains $\hat{\xi}_{t_0}$, the estimate of the state at time $t_0$, solves \eqref{eq:online_alloc}, allocates the network resources as $\textstyle{a_{t_0} = a^*_{t_0}}$, at time $t_0$, where $\bm{a}^*$ is the optimal solution, and repeats the process at the next time step. 
It is also possible to provide an ancillary network controller, or a baseline allocation scheme, based on either \eqref{eq:static_alloc} or \eqref{eq:dynamic_alloc}. In this case, the dynamic allocation scheme only accounts for online adjustments on the allocation in each time step, with the knowledge of the baseline allocation scheme.
By solving the problem \eqref{eq:online_alloc}, the online dynamic allocation scheme allocates network resources according to the requirements given by the DTs in terms of desired network allocation to provide the quality-of-service required to achieve the desired quality-of-control. 

\subsection{Event-Triggered Dynamic Allocation}

Solving \eqref{eq:online_alloc} in an online receding horizon fashion and updating the network allocation of resources may not be favorable in practical scenarios, due to computational or communication overhead, or due to complications with frequent allocation changes in the network.
We propose an event-triggered dynamic allocation scheme, which may be more suitable in practical applications.
Here, the network manager computes a re-allocation only when an event is triggered in the plant floor. 

Consider the plant floor and the physical systems in Figure~\ref{fig:setting}. Let the physical systems/components be indexed by $\textstyle{i\in\{1,\ldots,n\}}$. 
The DTs continuously monitor the output and performance of each physical component, and can identify  discrepancies due to either disturbances in the physical system or a mismatch between the required and allocated network resources. We model such performance discrepancy with a \textit{regret} function associated with each physical system:
\begin{equation}\label{eq:regret_func}
    R_{i}(T|\tau)=\sum_{t=\tau}^{\tau + T}\left(p_{i}(t) - \hat{p}_{i}(t)\right),
\end{equation}
where $\hat{p}_{i}(t)$ denotes the performance of system/component $i$ under the DT-requested resource allocation, and $p_{i}(t)$ is the actual measured performance of the system/component at time $t$. 
The metric $R_{i}(T|\tau)$ computes the cumulative performance discrepancy of the physical system/component $i$ starting from a time $\tau$ denoting the time of the last reallocation event and $T$ denoting the number of time-steps after $\tau$.

The regret functions can thus track real-time performance error of each component, and if the error exceeds some thresholds, adaptation or reconfiguration may be performed.  
The performance is considered as ``satisfactory'' as long as
\begin{equation}\label{eq:regret_bound}
    |R_i(T|\tau)|\leq \epsilon_i(\tau+T). 
\end{equation}
The thresholds $\epsilon_i(\tau+T)$ are determined \textit{a priori} depending on the controlled performance requirements and on the characteristics of the physical system $i$. Hence they may change from one physical system to the other and between different control tasks. Additionally, note that since \eqref{eq:regret_func} grows with the number of time steps, the bound $\epsilon_i(
tau+T)$ may also be set to increase as a function of time index to consider this growth, depending on the application setting.
The incremental difference $\textstyle{\left(p_{i}(t) - \hat{p}_{i}(t)\right)}$ in \eqref{eq:regret_func} may be negative if more than required network resources are allocated to a system. 
The threshold in \eqref{eq:regret_bound} penalizes such excessive allocation as the resources may be given to other systems in the network instead. The regret function \eqref{eq:regret_func} and the threshold \eqref{eq:regret_bound} may be adjusted based on the specific application of interest to reflect the satisfactory allocation scenarios.

The proposed scheme utilizes the regret bound \eqref{eq:regret_bound} to trigger a dynamic allocation event, which solves the following optimization problem
\begin{subequations}
\label{eq:event_alloc}
\begin{align}
	\min\limits_{a}~& \mathcal{I}(\bm{\gamma}) + \sum_{t = t_0}^{t_0+N_e} \ell(\xi_t,r_t,a), \\
	\text{s.t.:}~& H(\bm{r}) a \leq  \bm{b}(\bm{r})+ \bm{\gamma},\quad \xi_{t_0} = \hat{\xi} \\
	   ~&\xi_{t+1} = A \xi_t + B a,~ t = t_0,\ldots, t_0 +N_e-1, \label{eq:nw_dyn2}
\end{align}
\end{subequations}
where the decision variable $a$ now represents the fixed network allocation for the prediction horizon of $N_e$, since the network allocation is fixed and only updated after an event, i.e., $\textstyle{a_t = a}$ for $\textstyle{t = t_0, \ldots, t_0+N_e}$. Similar to the problem \eqref{eq:online_alloc}, the prediction of allocation requirements $\bm{r}$ and corresponding parameters are utilized as part of the constraints in $H(\bm{r})$ and $\bm{b}(\bm{r})$.

The horizon $N_e$ is an estimate for the timing of the next event. One way to obtain this estimate is through historical data. Let $\tau_k$ denote the time for the $k$th event in the past. Then the average time interval between the past $m$ events is an approximation for expected $N_e$, so that we have $\textstyle{N_e = \lfloor1/m \sum_{k=1}^{m-1} \tau_{k+1} - \tau_k\rfloor }$, where $\textstyle{\lfloor \cdot \rfloor}$ denotes the floor operator.
Alternatively, the network manager may utilize the DTs to simulate what-if scenarios and perform a bisection search, or a maximum reallocation period can be set so that the network manager reallocates after the specified time or if a threshold-based event is triggered. 

The dynamic allocation event may be triggered by including an appropriate triggering function that monitors the regret bound, as expressed in \eqref{eq:regret_bound}, in the DTs. For example, a trigger may be set when the mean regret, or maximum regret of all assets on the plant floor is above a preset amount. Using the measure of regret, the proposed allocation scheme is able to make controller-aware decisions that would satisfy requirements that ensure a desired quality-of-control. A prominent direction for future work is to study the effects of the triggering functions on the regret of the physical assets in the industrial plant.

\subsection{The role of DTs}
The dynamic allocation methods require the knowledge of $\bm{r}$, as well as, $\hat{\xi}$.
While a centralized network controller may directly obtain these information from the controllers of the physical resources, this task may scale poorly in a large scale CPMS with various heterogeneous resources. To this end, structured centralized control schemes such as software defined control~\cite{lopez2018software} are proposed to develop unified communication interfaces between the heterogeneous resources in the plant floor, and the centralized controller. While such applications mitigate the overhead of communication protocols, the centralized controller still requires a run-time representation of the physical resources which is provided by the DTs~\cite{moyne2020requirements}.

While DT information is often utilized in a monitoring capacity in the literature, here we utilize the DTs as a distributed network of observers 
and utilize the DT monitoring data as inputs to the centralized controller for online network allocation.

\begin{figure}
	\centering
	\includegraphics[width=\columnwidth]{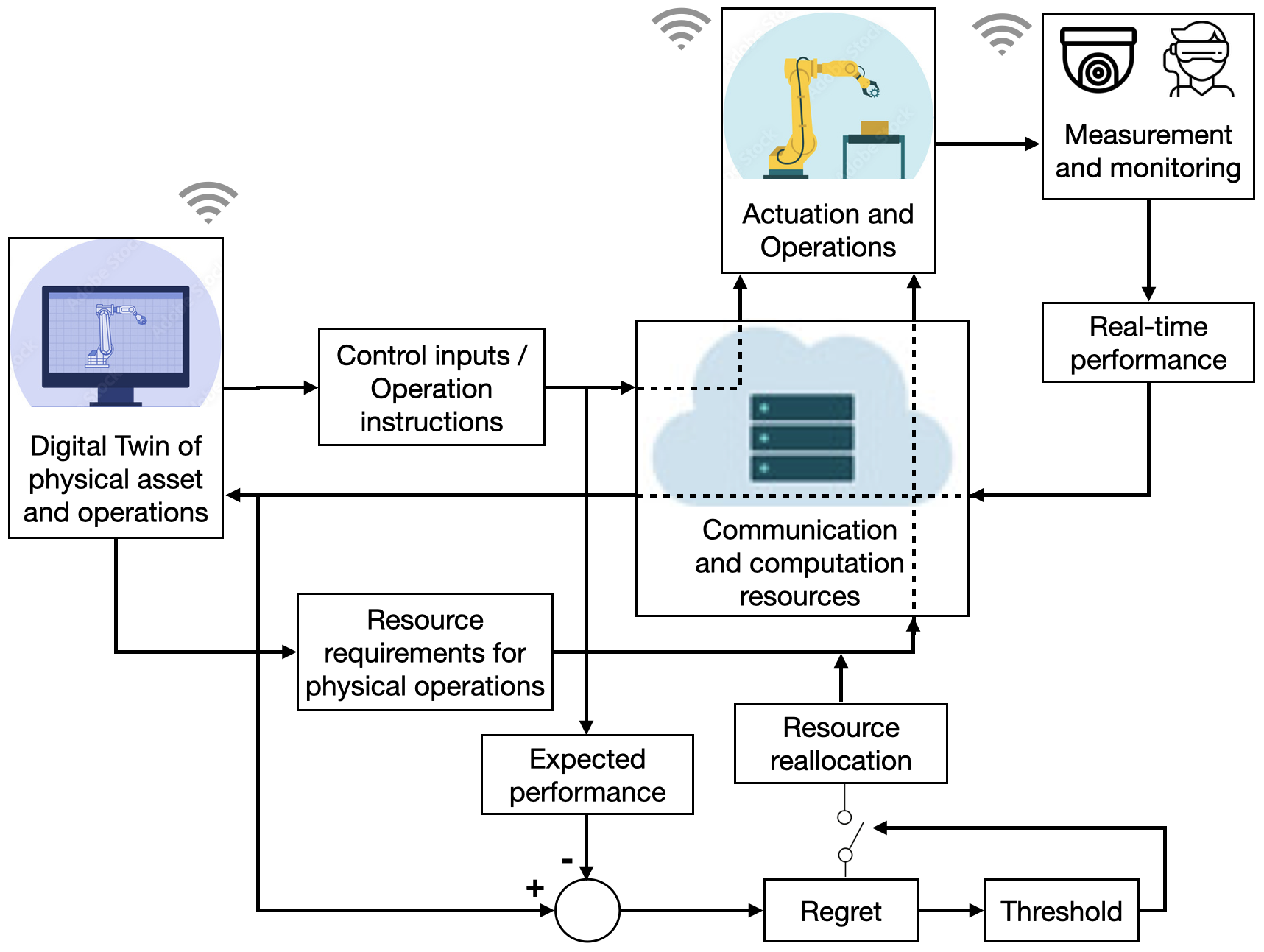}
	\caption{Schematic diagram of a CPMS scenario (from a single system perspective) with a network that supports the required communication and computation resources and with DTs with feedback to measure the regret and trigger resource reallocation, if necessary.}
	\label{fig:schematic_loop}
\end{figure}

Figure \ref{fig:schematic_loop} illustrates the proposed dynamic allocation concept from a single physical system's perspective. 
The measurement and monitoring data collected from the physical system are communicated to the central controller, through the unified communication interface (see Figure~\ref{fig:setting}).
DTs of the corresponding physical resources process this data and compute the control action via the computational resource allocated by the network manager, which is then applied on the physical system. 
Therefore, DTs provide a distributed network of computational intelligence that use common computational resources based on the allocation determined by the network manager. 
The performance regret of the process is also continuously monitored by the DTs, and in the case of event-triggered allocation, new allocations are computed when the regret metric is above a certain threshold. 

\section{Simulation Study}
\label{SecIV:SA}
We demonstrate the proposed network controller on a simulated plant floor with interconnected resources. The plant consists of $n_r$ physical resources, all of which have their own DTs. The DTs monitor the process outputs of the physical resources and apply control inputs. The control inputs are computed using optimization-based control methods, where the objective is to minimize a control cost, given a model and constraints of the system.

\subsection{Simulation Setting}

For the case study, we implement a proximal gradient algorithm (PGA) to solve at each simulation time $t$ for each DT. The PGA is used for computing the control action for each physical system.
PGA is an effective algorithm for optimizing convex functions over convex input constraints. Given an initial $\textstyle{x_k = x_0}$, the algorithm performs the following iterates 
\begin{align}\label{eq:pga}
  x_{k+1} = \Pi_{\mathcal{X}}\left( x_k - \alpha \nabla f(x_k)  \right),
\end{align}
where $\mathcal{X}$ is a convex input constraint set, $\alpha$ is the step size, and $\nabla f(x_k)$ is the gradient of the objective function at point $x_k$. 
Here, the input constraints may represent actuator constraints, and the objective function may be a quadratic function on the model-based tracking error. 
Under suitable conditions on the step-size, convexity of $f$, and the optimal solution, the PGA converges to the unique minimizer $x^*$~\cite{chambolle2016introduction}. 
Then, the control law to actuate the physical asset at time $\tau$ can then be defined as 
\begin{align}
    u_{\tau} = \sigma(x_f),
\end{align}
where $\sigma$ defines a static control policy on the final iterate of the PGA $x_f$. If the PGA is performed until convergence, we have $\textstyle{x_f = x^*}$.
Suppose the function $f$ is $L$-smooth. Then, choosing a constant step-size of $\textstyle{\alpha \leq 1/L}$ provides the suboptimality bound
\begin{align}\label{eq:pga_cost_convergence}
    F(x_k) - F(x^*) \leq \frac{|| x_0 - x^*||^2 }{2\alpha k},
\end{align}
where $\textstyle{F(x) = f(x) + \iota_{\mathcal{X}}(x)}$, with $\iota_{\mathcal{X}}$ being the indicator function for the convex set $\mathcal{X}$. In practice, the algorithm is run until a $\delta$ convergence criteria is reached or for a predefined fixed number of iterations due to computation time or capacity limits. 
Given the diameter of the convex constraint set $\mathcal{X}$ and the step size $\alpha$, it is possible to upper bound the right side of \eqref{eq:pga_cost_convergence} to a ball of $\delta$ as a function of $k$. Note that while this may not be a tight bound, it nevertheless provides a computable formal upper bound on how many iterations are needed to get the desired convergence ball. 
Let $k'$ be the required number of iterations to achieve the $\delta$ ball based on \eqref{eq:pga_cost_convergence}.
Note that since different resources may have different controllers and requirements, $f$, $\mathcal{X}$, and therefore, $k'$ might differ between resources and DTs.

In our setting, the computations of each DT are done over a networked resource which is available to the extent allowed by a network manager.
To run the PGA algorithm until convergence (e.g., for $k'$ steps), each DT requires a certain computational resource to perform the PGA updates.
The values of $k'$ and a minimum acceptable lower-bound $k_{\ell}$ is shared with the network manager in each time step.

To demonstrate the performance of the proposed network allocation schemes, we compare four scenarios, increasing in complexity, and provide results in the following section.
\begin{enumerate}[label=($C$\arabic*), leftmargin=*]
	\item \label{c:eql} Network resources allocated equally
	\item \label{c:sta} Network resources allocated according to policy \eqref{eq:static_alloc} with expected network requirements
	\item \label{c:tri} Event-triggered allocation scheme 
	\item \label{c:onl} Network resources allocated according to the online dynamic allocation, solving problem \eqref{eq:online_alloc} in receding horizon.
\end{enumerate}

The simulation is implemented in Matlab. The physical resources have their corresponding DTs that compute their network requirements, predict future network requirements, and compute control actions based on the PGA algorithm. All DTs provide the requirement prediction vectors to the network manager and the network manager provides the network allocation through a publish-subscribe channel in the simulation. There are 20 physical resources in the network and the simulation takes 100 time steps. The network requirement for all agents is stationary for the first 10 time steps, and updated with bounded random integers in future times. 
The maximum allowable deviation from the requested network allocation is set at 10 for all resources, e.g., $\textstyle{k'-k_{\ell} = 10}$.
For simplicity, let $A = 0$ and $B = I$ for the simulation study, so that the state $\xi$ in the optimization problem represents the current allocation in the network.
Consequently, we set $\textstyle{N=1}$ for the online allocation scheme in \eqref{eq:online_alloc}.

\subsection{Results}

\begin{figure}[h]
	\centering
	\includegraphics[width=0.95\columnwidth]{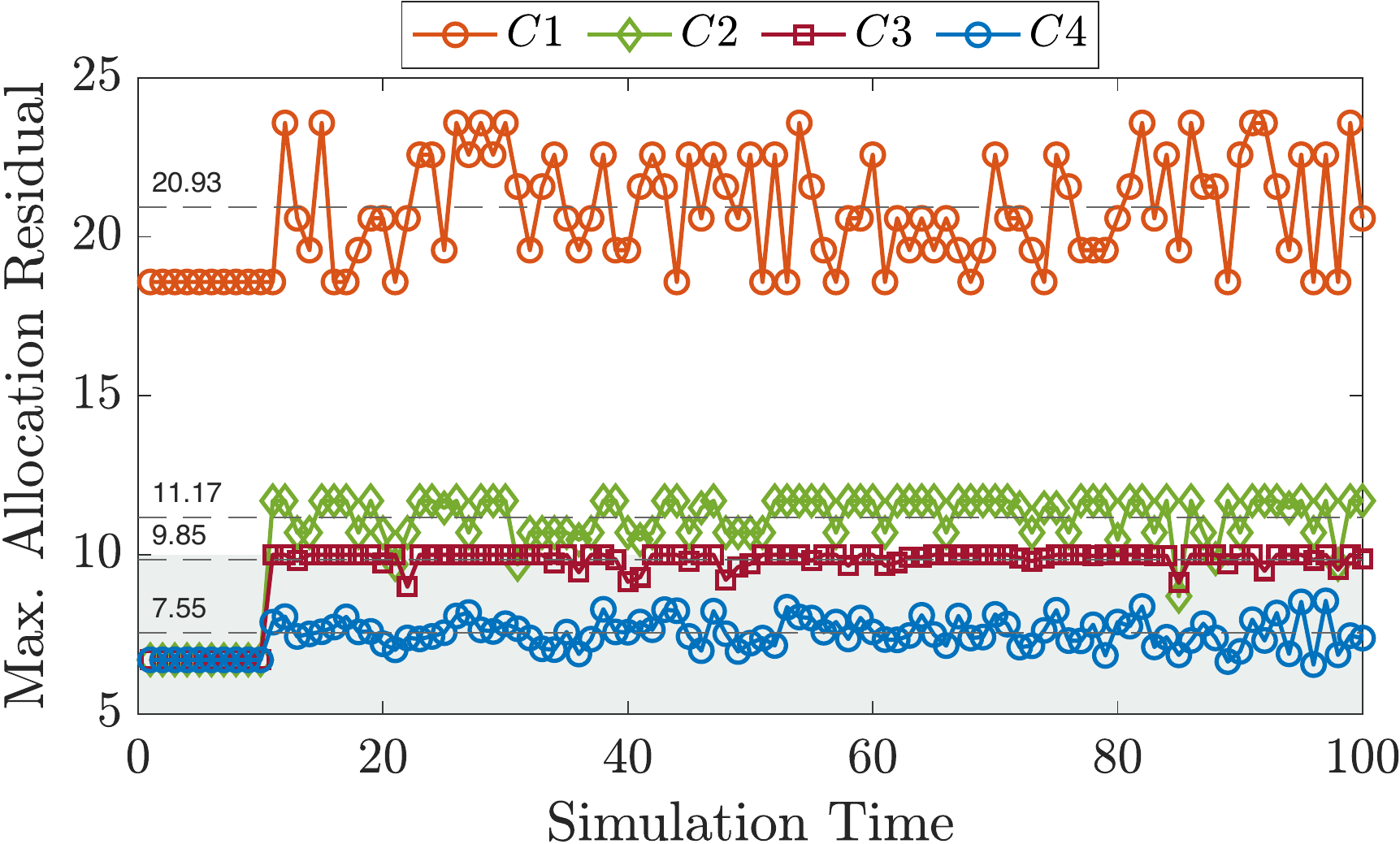}
	\caption{Comparison of all four network resource allocation scenarios compared in the case study. Mean values after time step 10 are shown with dashed lines and values on the left. The maximum allowable allocation residual is shown with colored background.
	}
	\label{fig:alloc_comparison}
\end{figure}

Figure~\ref{fig:alloc_comparison} illustrates the maximum network resource allocation residual at a given simulation tick, i.e., $||r_t - a_t||_{\infty}$ for the four simulation scenarios. Equal allocation in~\ref{c:eql} performs the worst as it does not account for the requirements provided by the DTs of the physical systems and the online changes. The scenarios \ref{c:sta}, \ref{c:tri}, and \ref{c:onl} have identical allocation residual at the beginning, representing the static allocation at the beginning. However, after time step 10, we see that the proposed online dynamic allocation-based controller outperforms all other methods.

Additionally, we see that the maximum resource residual is limited below 10 for the online allocation schemes \ref{c:tri} and \ref{c:onl}, shown with the colored background in Figure~\ref{fig:alloc_comparison}. Using the maximum allowable resource residual reported by the DTs, the proposed controller can achieve a noticeably better performance by using the run-time data to actively adjust the network resource allocation.
The online allocation \ref{c:onl} reallocates the resources in each time step, which may be undesirable in practical scenarios due to the high frequency of computing the allocation pattern that may itself require considerable computation power. 
The event-triggered scheme in \ref{c:tri} achieves comparable quality-of-service in the sense of keeping the maximum residual below the required level by reallocating only 15 times throughout the simulation.

\section{Conclusion and Future Work}
This paper proposes controller-aware network allocation schemes for efficient network resource management in CPMS settings.
The proposed framework demonstrates that the flexibility and robustness of novel networking technologies can be exploited together with the application of DT to efficiently respond to critical industrial operation requirements for ICPS and CPMS. 
We provide practical and deployable methods to continuously reallocate resources and controller performance-based metrics to perform reallocation based on events, and provide a simulation study to illustrate the efficiency of proposed dynamic resource allocation methods. 
Future work will consider further analysis on the effect of different triggering functions, the effect of additional network dynamics and inertia, and further analysis on the effect of control algorithms on the resource allocation architectures.

\label{SecV:Con}




%
%


\bibliographystyle{ieeetr}
\bibliography{Ref.bib}

\begin{thebibliography}{10}

\bibitem{8558534}
H.~Xu, W.~Yu, D.~Griffith, and N.~Golmie, ``A survey on industrial internet of
  things: A cyber-physical systems perspective,'' {\em IEEE Access}, vol.~6,
  pp.~78238--78259, 2018.

\bibitem{Allgoewer2019}
F.~Allgöwer, J.~Borges~de Sousa, J.~Kapinski, P.~Mosterman, J.~Oehlerking,
  P.~Panciatici, M.~Prandini, A.~Rajhans, P.~Tabuada, and P.~Wenzelburger,
  ``Position paper on the challenges posed by modern applications to
  cyber-physical systems theory,'' {\em Nonlinear Analysis: Hybrid Systems},
  vol.~34, pp.~147--165, 2019.

\bibitem{7778207}
S.~A. Seshia, S.~Hu, W.~Li, and Q.~Zhu, ``Design automation of cyber-physical
  systems: Challenges, advances, and opportunities,'' {\em IEEE Transactions on
  Computer-Aided Design of Integrated Circuits and Systems}, vol.~36, no.~9,
  pp.~1421--1434, 2017.

\bibitem{8840800}
L.~Monostori, B.~Kádár, T.~Bauernhansl, S.~Kondoh, S.~Kumara, G.~Reinhart,
  O.~Sauer, G.~Schuh, W.~Sihn, and K.~Ueda, ``Cyber-physical systems in
  manufacturing,'' {\em CIRP Annals}, vol.~65, no.~2, pp.~621--641, 2016.

\bibitem{8403588}
S.~Gangakhedkar, H.~Cao, A.~R. Ali, K.~Ganesan, M.~Gharba, and J.~Eichinger,
  ``Use cases, requirements and challenges of 5{G} communication for industrial
  automation,'' in {\em 2018 IEEE International Conference on Communications
  Workshops}, pp.~1--6, 2018.

\bibitem{8612449}
T.~Taleb, I.~Afolabi, and M.~Bagaa, ``Orchestrating 5{G} network slices to
  support industrial internet and to shape next-generation smart factories,''
  {\em IEEE Network}, vol.~33, no.~4, pp.~146--154, 2019.

\bibitem{9200919}
C.~Papagianni, J.~Mangues-Bafalluy, P.~Bermudez, S.~Barmpounakis,
  D.~De~Vleeschauwer, J.~Brenes, E.~Zeydan, C.~Casetti, C.~Guimarães,
  P.~Murillo, A.~Garcia-Saavedra, D.~Corujo, and T.~Pepe, ``5growth:
  A{I}-driven 5{G} for automation in vertical industries,'' in {\em European
  Conference on Networks and Communications}, pp.~17--22, 2020.

\bibitem{8931325}
J.~Ordonez-Lucena, J.~F. Chavarria, L.~M. Contreras, and A.~Pastor, ``The use
  of 5{G} non-public networks to support {I}ndustry 4.0 scenarios,'' in {\em
  2019 IEEE Conference on Standards for Communications and Networking (CSCN)},
  pp.~1--7, 2019.

\bibitem{8853247}
J.~García-Morales, M.~C. Lucas-Estañ, and J.~Gozalvez, ``Latency-sensitive
  5{G} {RAN} slicing for {I}ndustry 4.0,'' {\em IEEE Access}, vol.~7,
  pp.~143139--143159, 2019.

\bibitem{Bonati2020}
L.~Bonati, M.~Polese, S.~D’Oro, S.~Basagni, and T.~Melodia, ``Open,
  programmable, and virtualized 5{G} networks: State-of-the-art and the road
  ahead,'' {\em Computer Networks}, vol.~182, 2020.

\bibitem{8660405}
Z.~Zhou, Y.~Guo, Y.~He, X.~Zhao, and W.~M. Bazzi, ``Access control and resource
  allocation for {M2M} communications in industrial automation,'' {\em IEEE
  Transactions on Industrial Informatics}, vol.~15, no.~5, pp.~3093--3103,
  2019.

\bibitem{8399557}
L.~Yin, J.~Luo, and H.~Luo, ``Tasks scheduling and resource allocation in fog
  computing based on containers for smart manufacturing,'' {\em IEEE
  Transactions on Industrial Informatics}, vol.~14, no.~10, pp.~4712--4721,
  2018.

\bibitem{59155}
F.~Schnicke, T.~Kuhn, and P.~O. Antonino, ``Enabling {I}ndustry 4.0
  service-oriented architecture through digital twins,'' in {\em European
  Conference on Software Architecture}, pp.~490--503, 2020.

\bibitem{qamsane2019unified}
Y.~Qamsane, C.-Y. Chen, E.~C. Balta, B.-C. Kao, S.~Mohan, J.~Moyne, D.~Tilbury,
  and K.~Barton, ``A unified digital twin framework for real-time monitoring
  and evaluation of smart manufacturing systems,'' in {\em IEEE 15th
  international conference on automation science and engineering (CASE)},
  pp.~1394--1401, IEEE, 2019.

\bibitem{kovalenko2022towards}
I.~Kovalenko, J.~Moyne, M.~Bi, E.~C. Balta, W.~Ma, Y.~Qamsane, X.~Zhu, Z.~M.
  Mao, D.~M. Tilbury, and K.~Barton, ``Towards an automated learning control
  architecture for cyber-physical manufacturing systems,'' {\em IEEE Access},
  2022.

\bibitem{lopez2018software}
F.~Lopez, Y.~Shao, Z.~M. Mao, J.~Moyne, K.~Barton, and D.~Tilbury, ``A
  software-defined framework for the integrated management of smart
  manufacturing systems,'' {\em Manufacturing Letters}, vol.~15, pp.~18--21,
  2018.

\bibitem{9422344}
X.~Li, A.~Garcia-Saavedra, X.~Costa-Perez, C.~J. Bernardos, C.~Guimarães,
  K.~Antevski, J.~Mangues-Bafalluy, J.~Baranda, E.~Zeydan, D.~Corujo,
  P.~Iovanna, G.~Landi, J.~Alonso, P.~Paixão, H.~Martins, M.~Lorenzo,
  J.~Ordonez-Lucena, and D.~R. López, ``5growth: An end-to-end service
  platform for automated deployment and management of vertical services over
  5{G} networks,'' {\em IEEE Communications Magazine}, vol.~59, no.~3,
  pp.~84--90, 2021.

\bibitem{8311660}
M.~C. Lucas-Estañ, T.~P. Raptis, M.~Sepulcre, A.~Passarella, C.~Regueiro, and
  O.~Lazaro, ``A software defined hierarchical communication and data
  management architecture for industry 4.0,'' in {\em 2018 14th Annual
  Conference on Wireless On-demand Network Systems and Services (WONS)},
  pp.~37--44, 2018.

\bibitem{8553664}
A.~Ksentini, P.~A. Frangoudis, A.~PC, and N.~Nikaein, ``Providing low latency
  guarantees for slicing-ready 5{G} systems via two-level {MAC} scheduling,''
  {\em IEEE Network}, vol.~32, no.~6, pp.~116--123, 2018.

\bibitem{moyne2020requirements}
J.~Moyne, Y.~Qamsane, E.~C. Balta, I.~Kovalenko, J.~Faris, K.~Barton, and D.~M.
  Tilbury, ``A requirements driven digital twin framework: Specification and
  opportunities,'' {\em IEEE Access}, vol.~8, pp.~107781--107801, 2020.

\bibitem{7976223}
J.~Vachálek, L.~Bartalský, O.~Rovný, D.~Šišmišová, M.~Morháč, and
  M.~Lokšík, ``The digital twin of an industrial production line within the
  {I}ndustry 4.0 concept,'' in {\em 2017 21st International Conference on
  Process Control (PC)}, pp.~258--262, 2017.

\bibitem{langson2004robust}
W.~Langson, I.~Chryssochoos, S.~Rakovi{\'c}, and D.~Q. Mayne, ``Robust model
  predictive control using tubes,'' {\em Automatica}, vol.~40, no.~1,
  pp.~125--133, 2004.

\bibitem{mesbah2016stochastic}
A.~Mesbah, ``Stochastic model predictive control: An overview and perspectives
  for future research,'' {\em IEEE Control Systems Magazine}, vol.~36, no.~6,
  pp.~30--44, 2016.

\bibitem{chambolle2016introduction}
A.~Chambolle and T.~Pock, ``An introduction to continuous optimization for
  imaging,'' {\em Acta Numerica}, vol.~25, pp.~161--319, 2016.

\end{thebibliography}

\end{document}